# Market Coupling as the Universal Algorithm to Assess Zonal Divisions

Grzegorz Oryńczak, Marcin Jakubek, Karol Wawrzyniak, Michał Kłos
National Centre for Nuclear Research, Świerk Computing Centre
Otwock-Świerk, Poland
Karol.Wawrzyniak@fuw.edu.pl

*Abstract*— Adopting a zonal structure of electricity market requires specification of zones' borders. In this paper we use social welfare as the measure to assess quality of various zonal divisions. The social welfare is calculated by Market Coupling algorithm. The analyzed divisions are found by the usage of extended Locational Marginal Prices (LMP) methodology presented in paper [1], which takes into account variable weather conditions. The offered method of assessment of a proposed division of market into zones is however not limited to LMP approach but can evaluate the social welfare of divisions obtained by any methodology.

## I. INTRODUCTION

The energy market in Europe is undergoing a process of transformation aimed at integration of national markets and making better use of renewable generation sources. The market structure used in many countries, mostly due to historical reasons, is the uniform pricing, in which there is a single price of energy set on a national market for each hour of a day. In spite of its apparent simplicity, such an approach has serious disadvantages. The equilibrium set on the market does not take into account safety requirements of the grid. Hence, (i) the single-price equilibrium set on the market (energy exchange) is frequently unfeasible, (ii) the system operator has to perform costly readjustments, (iii) costs of supplying the energy differ between locations, but they are not covered where they arise.[1] With introduction of other forms of market, congestion costs are mitigated and the price on the market reflects the true costs of supplying energy to different locations in a more adequate way.

Hitherto, the explicit type of the future pan-European energy market remains an open question as the Third Energy Package does not specify the design precisely, but the two most popular approaches towards which national markets evolve are nodal and zonal pricing. The nodal pricing model is currently used in, among others, the US and Russia. Zonal pricing has been introduced in the Nordic countries as well as in Great Britain. This type of pricing is gaining in popularity across the Europe, and will be the main subject of this paper. Zonal market, which can be thought of as a compromise between simplicity of uniform structure and accuracy of nodal one, introduces differentiation of prices between regions with distinct costs of supplying energy, but it maintains the transparency which the nodal market lacks. The grid is divided into geographical regions (zones), each having a separate market for the energy with possibly different price. Market Coupling (MC) algorithm is used to control inter-zonal power flows and to calculate prices in zones given those flows. This way, under presumption that the zones were chosen so that frequently congested lines are on their borders, the equilibrium on zonal markets will take into account transfer limits on those critical lines. The need for additional congestion management is thus minimized, with most of the task being performed by the MC mechanism. Of course, small adjustments of equilibrium to satisfy limits on intra-zonal lines might be necessary, but they are expected to be less costly than the adjustments on a corresponding uniform market.

Still, there is no consensus in the literature with respect to methodology of identification of optimal zones' number and their borders. The existing methods are mostly based on two-stage approach – assignment of some specific values to each of the nodes and division of the system into regions by clustering the nodes over those parameters. Among existing methods, we can distinguish two popular approaches for choosing the values characterizing nodes.

The first approach is based on *nodal prices*, called also Locational Marginal Prices (LMPs) [2,3]. Nodal price represents the local value of energy, i.e. cost of supplying extra 1 MW of energy to a particular node - physical point in the transmission system, where the energy can be injected by generators or withdrawn by loads. This price consists of the cost of producing energy used at a given node and the cost of delivering it there taking into account congestion. Therefore LMPs separate locations into higher and lower price areas if congestion occurs between them. The second approach is based on Power Transfer Distribution Factors (PTDFs). The procedure starts from identification of congested lines, for which PTDFs are then calculated. The distribution factors reflect the influence of unit nodal injections on power flow along transmission lines, thus grouping the nodes characterized by similar factors into one zone defines a region of desirably similar sensitivity to congestions [4].

The main issue that has to be addressed concerning the ultimate choice of zonal configuration is the inconsistency in determinants used to separate the nodes and in the criteria used to evaluate constructed partitions. The two aforementioned methods derive divisions from a reasonable assumption that congested lines should become inter-zonal links, however, the actual shape of borderlines is based on

---

[1] For example, in Poland in 2011 the cost of the balancing market readjustments amounted to more than 3% (>250 million EUR) of the overall costs of production (*source: URE/ARE S.A.*).

This work was supported by the EU and MSHE grant nr POIG.02.03.00-00-013/09. The use of the CIS computer cluster at NCBJ is gratefully acknowledged.

several different premises used to (i) label each of nodes with a value (e.g. nodal price) or a set of values (e.g. PTDFs) and (ii) to group the nodes into geographic areas using a clustering technique.

Since the methods of division assign the values basing on data derived from nodal market structure (e.g. LMPs), there is a justified need for the assessment which evaluates a newly defined zonal architecture. We can broadly divide evaluation criteria into those based on safety of the network, which include, for example, predictability of the flows on intra-zonal lines, and those based on economic (market) measures, like market liquidity, concentration or social welfare (suggested by [7] among the others). In this article we calculate and compare social welfare (SW), defined as the sum of producers' and consumers' surplus in the equilibrium of supply and demand on the zonal market. Specifically, we use mechanism of Flow Based Market Coupling (FB MC) to determine the equilibria on each of the zonal markets, taking into account the limits of power flows between them.[2]

As a test case of divisions, we use the results of LMP methodology presented in [1], which takes into account variable weather conditions and is applied to a simplified model of Polish grid, and derive the appropriate welfare measures. The exact methodology of Market Coupling and calculation of social welfare is presented in the next section. In Sec. III we describe the model of network used as the test case. The results derived on it are presented and discussed in Sec. IV. In Sec. V we conclude and present directions for future work.

## II. THE METHODOLOGY

The zonal energy market can be represented as a set of energy exchanges, each governing the trade between generators and consumers of energy located in a particular geographic area. Energy transfers between zones are allowed and are governed by MC mechanism, which takes into account the constraints of the inter-zonal transmission lines. In order to determine safe supply-demand equilibria on each of energy exchanges, the MC mechanism must determine how the realization of buy/sell bids translate into (i) power injections/withdrawals in the nodes of the grid and into (ii) flows on inter-zonal lines. In doing so, the aim of the MC algorithm is to maximize the social welfare (consumer surplus + producer surplus + congestion rent) while keeping the flows in safety limits.

The mechanism of keeping the inter-zonal flows in safety limits can be of varying level of complexity. We use in our approach Flow Based MC, which takes advantage of the Power Transfer Distribution Factors (PTDF) matrix to determine how a particular injection/withdrawal influences power flows on inter-zonal lines. This approach is analogous to a Direct Current Power Flow (DC PF) model of the flows in the grid. As such, FB MC is a significant step-up in robustness compared to simpler MC mechanisms, for example the Available Transfer Capacity (ATC) MC, which limits only the aggregated flow between the zones, without calculating explicitly the flows on each of the inter-zonal lines.

---

[2] We use the term "Market Coupling" specifically as the mechanism governing the exchange between market zones which are managed by one system operator (for example, the case of a national market divided into zones, or a common zonal market spanning across more than one country). Governance of the flows on the borders of two (or more) not integrated markets is not studied in this paper.

We based our implementation of FB MC on the description of the COSMOS/Euphemia algorithm [5,8], which was derived for the CWE (Central Western Europe) energy market.

### A. Input of Offers

The main input data for the MC algorithm are the buy/sell offers of the energy consumers/generators. An $n$-th offer, $n = 1,…, N$, is characterized by affiliation to a zone $j$, $j = 1,…, J$, which we denote by $n \in Z_j$ and which is derived from the physical location of the node where the energy will be injected/withdrawn, and by the volume $q_n$ to be traded, which is coded as a positive amount for a sell bid, and negative for a buy bid. The algorithm allows partial realization of an offer and also the use of "triangle" offers, in which the offer price either decreases (for a buy offer) or increases (for a sell offer) linearly on the interval $[0, |q_n|]$, thus the offer is characterized by two prices, $P_n^0$ and $P_n^1$, that is, $P_n^0 \geq P_n^1$ for a buy offer, $P_n^0 \leq P_n^1$ for a sell offer. In the equilibrium found by the algorithm, each of the offers can be either accepted, accepted partially or not accepted at all, which will be coded by a coefficient $A_n \in [0,1]$. Thus, an offer $n$ accepted in percentage $A_n$ is connected with an injection/withdrawal of power in the amount of $A_n q_n$, with the highest price in case of "triangle" sell bid (lowest in case of buy bid) denoted by $\overline{P}_n = P_n^0 + (P_n^1 - P_n^0)A_n$.

### B. Flow Calculation

To calculate how the realization of offers affect inter-zonal flows for a given vector of zonal injections $Q = (Q_1,…,Q_J)^T$ (the aggregated injections/withdrawals representing accepted offers in all nodes in a given zone), which coordinates are given by

$$Q_j = \sum_{n \in Z_j} A_n q_n , \qquad (1)$$

we use the nodal PTDF matrix (*nPTDF*), derived from the model of the grid, and the Generation Shift Key (*GSK*) matrix [6] to obtain the flows along $K$ inter-zonal lines (vector $\tilde{Q} = (\tilde{Q}_1,…,\tilde{Q}_K)^T$) as $\tilde{Q} = nPTDF\ GSK\ Q$.

To construct the GSK matrix we firstly run the MC algorithm with the flows calculated without the use of *GSK* as $\tilde{Q} = R\ nPTDF\ Q$, where R matrix selects the flows along $K$ inter-zonal lines. The load/generation equilibrium, found for such constraints, is used to calculate the *GSK* matrix, which is then treated as input to the "proper" MC FB algorithm's run.

### C. Objective Function and Constraints

Maximization of the social welfare is equivalent to the optimization problem

$$\max \sum_{n=1}^{N} -q_n A_n \frac{\overline{P}_n + P_n^0}{2}. \qquad (2)$$

The safety limits on the *K* inter-zonal lines are described by their capacities $C_k$, $k \in 1,...,K$, with vector *C* defined as $C = (C_1, ..., C_K)^T$. For the vector of zonal injections *Q* given by (1), the safety constraints on flows in inter-zonal lines, are characterized by the condition $|\tilde{Q}| \leq C$, which translates to

$$|nPTDF\ GSK\ Q| \leq C. \qquad (3)$$

Lastly, we add the balance condition, which states that the sum of energy bought on a market of zone $Z_j$ and imported to the zone $Z_j$ must be equal to the energy sold on this market and exported from zone $Z_j$, that is,

$$\sum_{n \in Z_j} A_n q_n = \sum_{l \in from(Z_j)} \tilde{Q}_l - \sum_{l \in to(Z_j)} \tilde{Q}_l, \qquad (4)$$

where by *from*($Z_j$) and *to*($Z_j$) we denote the sets of the inter-zonal lines along which the energy flows are, respectively, withdrawing power from and injecting power into zone $Z_j$. The optimization problem described by objective function (2) and constraints (3) and (4) is then input to the IBM CPLEX mixed integer quadratic solver, and the vector of offer acceptance levels $A = (A_1,...,A_N)$ is derived.

### D. Market Clearing Prices

When the acceptance levels *A* are found, we can identify the Market Clearing Prices (MCPs) for each of the zone $Z_j$, namely $MCP_j$, $j = 1,..., J$. In general, a clearing price on a market is defined as any price for which the aggregated supply and demand (taking into account the import/export flows) is in equilibrium. Since such price might be not determined exactly (there might exist a range of prices satisfying the equilibrium condition), we define $MCP_j$ in the following way, which was chosen to accommodate the non-elastic demand assumption (cf. section III): (i) if demand in zone $Z_j$ was satisfied (for all buy offers in zone $Z_j$ we have $A_n = 1$), then we take as $MCP_j$ the highest price of a sell offer accepted at or imported to $Z_j$; (ii) if demand in zone $Z_j$ was not satisfied completely (there exist a buy offer in zone $Z_j$ such that $A_j < 1$), then we take as $MCP_j$ the common price of buy offers, $\overline{P}$, as defined in section III.

### E. Social Welfare

The social welfare in the market equilibrium found by the above procedure is equal to

$$SW = \sum_{j=1}^{J} \left[ \sum_{n \in Z_j} A_n q_n \left( MCP_j - \frac{\overline{P}_n + P_n^0}{2} \right) \right] \\ + \sum_{j=1}^{J} \sum_{k>1}^{J} (MCP_j - MCP_k) f_{jk}, \qquad (5)$$

where second double sum is the overall congestion rent of the system operator ($f_{jk}$ indicates power transfer between adjacent zones *j* and *k*).

### F. Redispatch Costs

Since the MC mechanism controls the congestion of only inter-zonal lines, there is a possibility that system operator will have to correct the generation profile set on the market in order to avoid congestion on intra-zonal lines. This process of redispatch generates additional costs since some of the generation has to be shifted from the cheaper to more expensive producers. Thus, in order to better reflect the true costs of supplying energy in our social welfare measure, we correct the amount (5) by estimator of the redispatch costs. To this end, on the load/generation profile acquired as a solution from MC, we run Power Flow algorithm to calculate flows on intra-zonal lines and we compare them with the lines' capacities. If a line *l* in zone $Z_j$ is congested by amount of $o_l$ MW, we add to the cost of redispatch the amount $o_l P_j^{max}$, where $P_j^{max}$ is the highest cost of generation in zone $Z_j$, to obtain an upper bound on the redispatch costs.

### III. THE TEST CASE

To test our approach on an exemplary case which can be treated as a relatively close representation of a real energy network, we used the data on the Polish grid based on the case file included in MATPOWER distribution [6]. This case represents the Polish 400, 220 and 110 kV network during winter 1999-2000 peak conditions.

The system consists of 327 power generators, 2383 buses, and 2896 interconnecting branches. The quality of the data is unknown but the fact that the assumed costs of generation are linear gives a premise that these data are of rather poor quality. Hence, the analysis should be treated as exemplary, with no constructive conclusions for the Polish system. We used this specific case since this is the only available one in which congestion exists under base case load. Additionally, we decreased capacity limits of two specific branches in order to obtain more pronounced influence of congestion on the MC solutions. A more detailed description of the data can be found in [1].

We divided the grid taking into account variable wind generation by using the method presented in [1]. We used three variations of our methodology, which reflect division for (i) no wind output, (ii) maximal wind output registered in the period between years 2007–12, and (iii) so called consensus clustering, which reflects aggregation of 722 different divisions made from various weather conditions into one. In each of the three variants we obtained divisions into 2, 3, 4 and 5 clusters, resulting in twelve divisions to be evaluated by the MC algorithm. Each of these 12 divisions was then tested in conditions reflecting average wind conditions in the period between years 2007–12.

Specifically, for each division we constructed offers of wind generators taking the estimated power output for average wind levels and offering it for a price equal to zero in order to secure the "sell" of the wind energy, which has priority over the conventional generation in the system. For conventional generators the energy available for sell, $q_n =$

$p_{max}$, where $p_{max}$ is the maximal output of the power plant, is offered at a constant price ($P_n^0 = P_n^1$), since the costs of generation in the used case file are linear. That is, we assume that the generators bid the available amount of energy by the marginal cost of production.

From the consumers' side, since only constant loads at each bus are available in the data, we assume that demand is perfectly inelastic, namely, that the loads at each bus are expected to be covered at any cost. To input such buy offers to the algorithm, for each nodal load we use an offer with $q_n$ equal to the negative of load, and with the price set at a level $P_n^0 = P_n^1 = \overline{P}$ common for all demand offers. In calculations presented below we have arbitrarily chosen $\overline{P} = 2000$ PLN, which is greater than any production cost in the data. Since we are interested not in the absolute level of social welfare for each proposed zonal division, but in the comparison of the social welfares obtained across the 12 different divisions, the choice of the demand bid price has no influence on the results, as long as this price is equal in every division case and is higher than all the sell offers' prices.

## IV. RESULTS & DISCUSSION

Since, as yet, there is no widely accepted methodology for choosing the exact number of zones into which market should be divided (although there have been some attempts, cf. [11]), in our study we assumed the range of the three aforementioned variants of divisions (no wind, maximal wind, consensus) to, respectively, 2, 3, 4 and 5 clusters, which yielded 12 division cases. Then we analyzed each of them separately and compared the results of social welfare in those 12 cases. In Tables 1&2 we characterize the results quantitatively, while Figure 1 delineates the geographical placement for divisions into 3 and 4 zones. Tab. 1 shows differences between SW, redispatch costs and SW corrected by redispatch levels calculated for zonal divisions with respect to single-zone market as a reference point.[3] In the case of single-zone market, SW is calculated from the MC solution in which there is only one zone, thus no congestion management is embedded in market mechanism and all is done by redispatch. SW in such case amounts to 47 470 970 PLN, redispatch costs: 116 589 PLN, SW corrected by redispatch: 47 354 381 PLN, marginal clearing price: 147 PLN.

The results summarized in Tab. 1 show that, as was expected, the (uncorrected) SW in the case of a single-zone market turned out to be the highest, since no congestion constraints are then put on the market solution. However, high redispatch costs associated with correction of this solution lead to the lowest corrected SW for single-zone market. The best division (with the highest SW corrected by redispatch) are related to 'max wind' variant for 3 clusters. In turn, the worst results are obtained for 2 clusters divisions in the 'consensus' and 'max wind' variants. One can notice that the corrected SW rises while redispatch costs drop with increasing number of clusters up to 4 (for 'no wind' and

'consensus' variants) or 3 (for 'max wind') and then both become relatively stable. This can be interpreted as a result of the most congested lines being taken into account as inter-zonal constraints into MC mechanism (instead of the costly redispatch) when the market is divided into 4 (or 3 in case of 'max wind') zones. Further increasing the number of zones does not lead to significant improvement.

| Number of zones / type of division | | Variant 1 no wind | Variant 2 consensus | Variant 3 max wind |
|---|---|---|---|---|
| 2 | SW | -1 585 | 0 | 0 |
| | redispatch | -7 144 | -3 558 | -3 558 |
| | SW corr | **5 559** | **3 558** | **3 558** |
| 3 | SW | -1 413 | -2 087 | -26 357 |
| | redispatch | -7 147 | -7 139 | -107 163 |
| | SW corr | **5 734** | **5 052** | **80 806** |
| 4 | SW | -30 943 | -30 130 | -27 234 |
| | redispatch | -90 622 | -103 082 | -107 132 |
| | SW corr | **59 679** | **72 952** | **79 898** |
| 5 | SW | -31 460 | -30 129 | -27 650 |
| | redispatch | -90 645 | -103 083 | -108 346 |
| | SW corr | **59 185** | **72 954** | **80 696** |

Table 1. Social welfare, redispatch costs and corrected SW in relation to single-zone market from for grid divisions into two, three, four and five zones done in 3 variants - no wind/consensus/maximal wind.

| | | # nodes | total power demand [GW] | # generators | total output [GW] | Market Clearing Price [PLN/MWh] |
|---|---|---|---|---|---|---|
| 2 zones | | 2325 | 23.0 | 357 | 24.1 | 147 |
| | | 2400 | 24.3 | 366 | 24.6 | 147 |
| | | 2401 | 24.3 | 366 | 24.6 | 147 |
| | | 97 | 1.6 | 9 | 0.5 | 2000 |
| | | 22 | 0.3 | 0 | 0 | 147 |
| | | 21 | 0.3 | 0 | 0 | 147 |
| 3 zones | yellow | 2325 | 23.0 | 357 | 24.1 | 147 |
| | | 2234 | 21.8 | 343 | 23.3 | 147 |
| | | 1592 | 15.3 | 218 | 18.9 | 129 |
| | green | 22 | 0.3 | 0 | 0 | 2000 |
| | | 22 | 0.3 | 0 | 0 | 2000 |
| | | 21 | 0.3 | 0 | 0 | 178 |
| | purple | 75 | 1.3 | 9 | 0.5 | 2000 |
| | | 166 | 2.5 | 23 | 1.3 | 2000 |
| | | 809 | 8.9 | 148 | 5.6 | 159 |
| 4 zones | yellow | 1477 | 14.8 | 255 | 12.1 | 152 |
| | | 1286 | 12.6 | 163 | 15.5 | 129 |
| | | 1592 | 15.3 | 218 | 18.9 | 129 |
| | grey | 848 | 8.1 | 102 | 11.9 | 129 |
| | | 948 | 9.2 | 180 | 7.8 | 159 |
| | | 734 | 7.6 | 139 | 5.1 | 159 |
| | purple | 75 | 1.3 | 9 | 0.5 | 2000 |
| | | 166 | 2.5 | 23 | 1.3 | 2000 |
| | | 75 | 1.3 | 9 | 0.5 | 2000 |
| | green | 22 | 0.3 | 0 | 0 | 2000 |
| | | 22 | 0.3 | 0 | 0 | 2000 |
| | | 21 | 0.3 | 0 | 0 | 2000 |

Table 2. Qualitative data for division of the Polish power grid into 2,3,4 zones. Each cell includes values for no-wind / consensus / max-wind variant in that vertical order. The colors reflect the areas depicted on Fig. 1.

---
[3] Namely, the values in Table 1 show the differences between appropriate levels for the solutions for market divided into 2, 3, 4, 5 zones and the levels for single-zone solution.

Looking at Tab. 2, one can notice that usually in a bigger zone the price is lower, in contrary to a smaller zone, where the final price is higher than the one for the single-zone market (147 PLN). Hence, when the bigger zone gets a lower price, then the total SW increases. (However, one must keep in mind that we do not include the redispatch costs in the prices calculated as in Section II.D, thus the link between the MCPs and SW corrected by the redispatch costs, which can be treated as the best estimator of market efficiency, is not straightforward.) Another reason for the SW divergence between the divisions is that demand in tiny zones, especially those not having their own generation, is sometimes not fully satisfied (thus MCPs there are often equal to $\overline{P} = 2000$ PLN). In sum, and in relation to SW levels in Tab. 1, one can see that the division with the highest corrected SW ('max wind,' 3 zones) is the one which has the smallest MCP in the biggest zone, while the demand is fully satisfied in all of the zones.

## V. FUTURE WORK

Among the issues that ought to be tackled in the future research we can distinguish two main subjects.

First, we acknowledge the need for improvement of the LMP-based clustering algorithm and development of other partitioning techniques that result in all of the zones being equipped with their own generation. The partitioning methods which produce bigger bidding areas are expected to eliminate the problem of exclusively external supply of energy. Also, the case of zones which overlap due to interfusion of the corridors formed along different types of transmission lines (e.g. 220 kV and 400 kV, cf. top right configuration on Fig. 1) remains unsolved.

Second, in the zonal approach all nodes in a specific zone are aggregated into one node. The influence of the power injected into this zone via transfer through the branches is estimated by zonal PTDFs (zlPTDF) matrices [6]. The calculation of zlPTDFs requires some assumption about ratios between generations/loads and net export which are expressed by GSK matrix. Thus, MC has to work with inflexible constraints given by GSK, where certain proportions between loads and generation are constant.

In other words, GSK has to be given as an input to MC. Thus, its value has to be guessed *a priori* before MC optimization starts. Then, in the optimization process, MC can select different combination of generations/loads than the one assumed in GSK, which subsequently forces incorrect evaluation of constraints.

Thus, as the consequence of the rough estimation of GSK matrices, the MC algorithm operates on unsound prerogatives. Both under- and overestimation of power flows along the transmission lines lead to suboptimal use of the infrastructure [6]. Thus, the process of deriving reliable GSK/zonal PTDFs is the central task for enhancement of zonal market stability and efficiency.

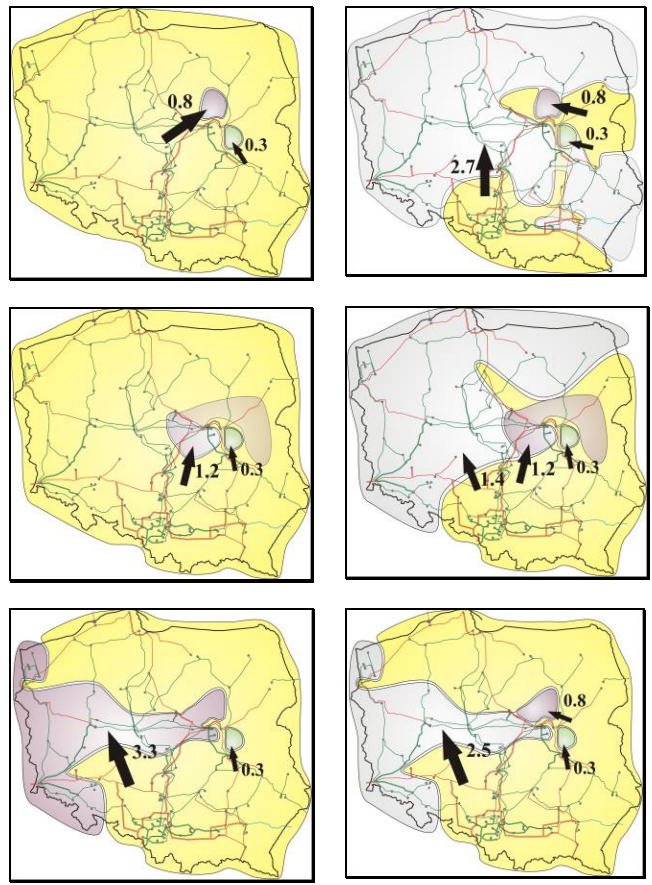

Figure 1. Polish grid division for three (left) and four (right) zones. Results for no-wind, consensus and maximal wind division variants are shown in the top, middle and bottom row, respectively. Arrows indicate direction and magnitude of energy transfers between zones in gigawatts.